\documentclass[sn-mathphys]{sn-jnl}
\usepackage{graphicx}

\theoremstyle{thmstyleone}%
\theoremstyle{thmstyletwo}%

\theoremstyle{thmstylethree}%

\begin{document}

\title[Synthetic Chiral Matter Based on CNTs]{Controlled synthetic chirality in macroscopic assemblies of carbon nanotubes}

\author[1,2,3]{Jacques Doumani}

\author[3]{Minhan Lou}

\author[4,5]{Oliver Dewey}

\author[6]{Nina Hong}

\author[3]{Jichao Fan}

\author[1,7]{Andrey Baydin}

\author[8]{Yohei Yomogida}

\author[8]{Kazuhiro Yanagi}

\author[4,5,7,9,10]{Matteo Pasquali}

\author[11]{Riichiro Saito$^{11}$}

\author[1,4,7,10,12]{Junichiro Kono}

\author*[3,4]{Weilu Gao}\email{weilu.gao@utah.edu}

\affil[1]{Department of Electrical and Computer Engineering, Rice University, Houston, TX, USA}

\affil[2]{Applied Physics Graduate Program, Smalley--Curl Institute, Rice University, Houston, TX, USA}

\affil[3]{Department of Electrical and Computer Engineering, The University of Utah, Salt Lake City, UT, USA}

\affil[4]{Carbon Hub, Rice University, Houston, TX, USA}

\affil[5]{Department of Chemical and Biomolecular Engineering, Rice University, Houston, TX, USA}

\affil[6]{J.A. Woollam Co., Inc., Lincoln, NE, USA}

\affil[7]{Smalley--Curl Institute, Rice University, Houston, TX, USA}

\affil[8]{Department of Physics, Tokyo Metropolitan University, Tokyo, Japan}

\affil[9]{Department of Chemistry, Rice University, Houston, TX, USA}

\affil[10]{Department of Materials Science and NanoEngineering, Rice University, Houston, TX, USA}

\affil[11]{Department of Physics, Tohoku University, Sendai, Japan}

\affil[12]{Department Physics and Astronomy, Rice University, Houston, TX, USA}

\newpage

\abstract{There is an emerging recognition that successful utilization of chiral degrees of freedom can bring new scientific and technological opportunities to diverse research areas. Hence, methods are being sought for creating artificial matter with controllable chirality in an uncomplicated and reproducible manner. Here, we report the development of two straightforward methods for fabricating wafer-scale chiral architectures of ordered carbon nanotubes (CNTs) with tunable and giant circular dichroism (CD). Both methods employ simple approaches, (i)~mechanical rotation and (ii)~twist-stacking, based on controlled vacuum filtration and do not involve any sophisticated nanofabrication processes. We used a racemic mixture of CNTs as the starting material, so the intrinsic chirality of chiral CNTs is not responsible for the observed chirality. In particular, by controlling the stacking angle and handedness in (ii), we were able to maximize the CD response and achieve a record-high deep-ultraviolet ellipticity of 40~$\pm$~1\,mdeg/nm. Our theoretical simulations using the transfer matrix method reproduce the salient features of the experimentally observed CD spectra and further predict that a film of twist-stacked CNTs with an optimized thickness will exhibit an ellipticity as high as 150\,mdeg/nm. The created wafer-scale objects represent a new class of synthetic chiral matter consisting of ordered quantum wires whose macroscopic properties are governed by nanoscopic electronic signatures such as van Hove singularities. These artificial structures with engineered chirality will not only provide playgrounds for uncovering new chiral phenomena but also open up new opportunities for developing high-performance chiral photonic and optoelectronic devices.}

\maketitle

\newpage

Chirality is a degree of freedom in objects with broken mirror symmetry, which ubiquitously exists in both natural and synthetic matter, ranging from molecules through crystals to metamaterials. There is much current interest in studying chiral effects in electronic and photonic processes in materials that can lead to new device concepts and implementations~\cite{AielloetAl22ACS,YangetAl21NRP,LodahletAl17Nature}. In particular, the interaction of chiral matter with circularly polarized light has profound implications and consequences in sensing~\cite{BarronEtAl2009}, plasmonics~\cite{HentscheletAl22SA}, cryptographic imaging and communication~\cite{YoonEtAl2018N,LiuEtAl2019N,HanEtAl2020AFM}, and quantum optics~\cite{LodahletAl17Nature,HubenerEtAl2021NM,ChenEtAl2021N}. 

The key enabler of these diverse applications is a versatile macroscopic chiral platform with large and controllable chiroptical properties. Chiral metamaterials, consisting of periodic artificial structures with symmetry breaking, can have significantly boosted chiroptical effects, compared to natural molecules, through resonant enhancement~\cite{WangEtAl2016N}. However, metamaterials usually require sophisticated nanofabrication facilities and intricate processes, especially for short-wavelength applications that necessitate ultrasmall feature sizes. Nanomaterials and their artificial architectures have recently emerged as new chiral platforms. For example, chirally stacked multiple layers of graphene have displayed circular dichroism (CD), which is defined as the differential absorption of left and right circularly polarized light~\cite{KimEtAl2016NN}. New methods that can create artificial matter with strong and controllable chirality without involving complicated procedures are being sought.

Here, we present two extraordinarily simple and scalable procedures, both based on the controlled vacuum filtration (CVF) method~\cite{HeetAl16NN,GaoKono19RSOS}, for preparing wafer-scale chiral matter consisting of ordered carbon nanotubes (CNTs). The first involves mechanical rotation during CVF, whereas the second uses twist-stacking of aligned CNT films produced by CVF. In both approaches, we employed a racemic mixture of CNTs, and hence, the chirality manifested by the produced architectures is not due to the intrinsic chirality of chiral CNTs. CD spectra exhibited spectral peaks reflecting the electronic structure of the underlying individual CNTs, which are quantum wires with one-dimensional van Hove singularities. Through precise control of stacking angle and handedness in the twist-stacking approach, we obtained a record-high ellipticity of 40~$\pm$~1\,mdeg/nm in the deep-ultraviolet (DUV) range (200--280\,nm). We performed electromagnetic simulations based on the transfer matrix method, which reproduced all observed CD and absorption spectra while simultaneously determining the dielectric functions of the aligned CNT films. We further predict that a film of twist-stacked CNTs with an optimized thickness will exhibit an ellipticity as high as 150\,mdeg/nm. This CNT-based chiral platform will open up new opportunities not only for next-generation photonic and optoelectronic devices, such as chiral (quantum) optical emitters~\cite{HeEtAl2017NP}, chiral optical sensors, and chiral photodetectors, but also for exploration of new phenomena, including non-optical effects, for broader applications.

\begin{figure}[hbt]
    \centering
    \includegraphics[width=1.0\textwidth]{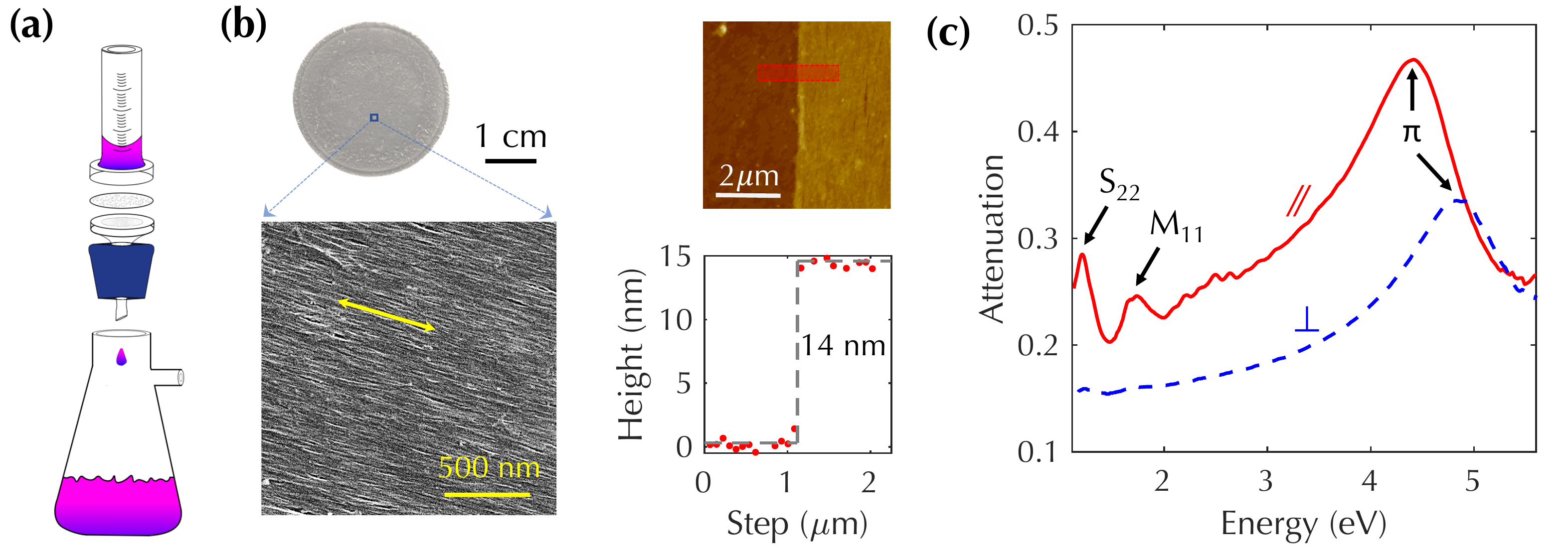}
    \caption{\textbf{Aligned CNT films produced by CVF and their strongly polarization-dependent optical response}. (a)~Schematic of the vacuum filtration system. (b)~Representative photograph (left top), scanning electron microscopy image (left bottom), atomic force image (right top), and a height profile (right bottom) of an obtained aligned CNT film. (c)~Polarization-dependent optical absorption spectra for an aligned CNT film from the near-infrared to the DUV.}
    \label{fig:fig1}
\end{figure}

Figure~\ref{fig:fig1}a schematically shows the vacuum filtration system we used, where individually suspended CNTs in water spontaneously assemble to form a wafer-scale, highly aligned, and densely packed film on the filter membrane in a well-controlled manner~\cite{HeetAl16NN,GaoKono19RSOS,KomatsuetAl20NL}; see Methods for more details. Figure~\ref{fig:fig1}b shows a photograph, a scanning electron microscopy image, and an atomic force microscopy image of a typically obtained aligned CNT film. The original aqueous suspension contained a racemic mixture of CNTs of both semiconducting and metallic types, showing zero CD (see Supplementary Fig.\,1). The experimentally measured quantity of CD is the ellipticity ($\psi$), which is defined as $\psi \equiv \arctan\left\{ (E_\mathrm{l} - E_\mathrm{r})/(E_\mathrm{l} + E_\mathrm{r}) \right \}$, expressed in units of mdeg. Here, $E_\mathrm{l}$ and $E_\mathrm{r}$ are the magnitudes of the left- and right-circular components, respectively, of the output electric field when the input beam that enters the material is linearly polarized~\cite{NordenEtAl2010}. We further normalized the ellipticity by the film thickness for the purpose of comparing different samples, so $\psi$ is given in units of mdeg/nm. Figure~\ref{fig:fig1}c shows absorption spectra for an aligned CNT film on a fused silica substrate when the incident light polarization is parallel and perpendicular to the CNT alignment direction. In addition to the near-band-edge interband transitions -- S$_{22}$ of semiconducting CNTs at 1.2\,eV and M$_{11}$ of metallic CNTs at 1.7\,eV -- there is a strong DUV response at 4.4\,eV (or 281\,nm) for parallel polarization and 4.87\,eV (or 255\,nm) for perpendicular polarization~\cite{MurakamiEtAl2005PRL}. This DUV spectral feature (labeled `$\pi$') originates from interband transitions associated with the van Hove singularity at the $M$ point of the graphene Brillouin zone~\cite{TakagiEtAl2009PR,MakEtAl2011PRL}.

\begin{figure}[hbt]
    \centering
    \includegraphics[width=1.0\textwidth]{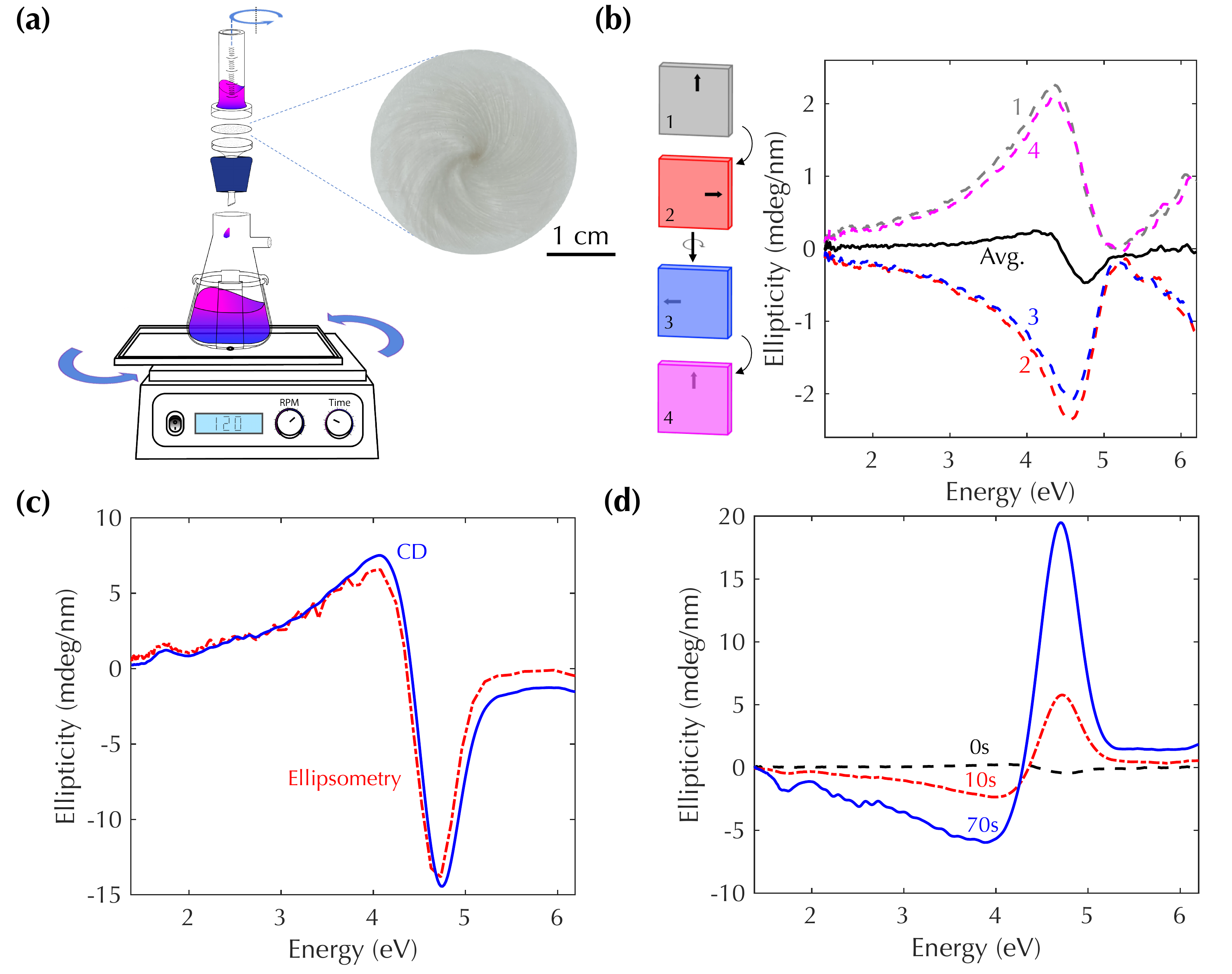}
    \caption{\textbf{Mechanical-rotation-assisted CVF for creating synthetic chiral matter based on ordered carbon nanotubes.} (a)~Schematic of the mechanical-rotation-assisted vacuum filtration setup. The standard filtration system was mounted on an orbital mechanical shaker, and the rotational motion produced a twisted CNT thin-film architecture. (b)~CD spectra for a highly aligned CNT film under four measurement configurations and their average. (c)~CD spectra for a twisted CNT architecture measured using a CD spectrometer with the four-configuration approach and spectroscopic ellipsometry. (d)~CD spectra of the twisted CNT films prepared with 0\,s, 10\,s, and 70\,s rotation time periods.}
    \label{fig:fig2}
\end{figure}

We first describe the first approach -- \emph{in~situ} mechanical rotation during vacuum filtration -- to convert spontaneous alignment into a wafer-scale twisted structure. As shown in Fig.\,\ref{fig:fig2}a, the filtration system was placed on an orbital mechanical shaker, which rotated the whole system on a horizontal plane around a vertical axis. Rotational shaking was applied for a short period of time when the liquid level inside the funnel was near the top of the filter membrane. The rotation speed and displacement were fixed; see Methods for more details, Supplementary Fig.\,2 for an illustration, and Supplementary Video 1 for an experimental demonstration. The applied rotational motion caused suspended CNTs to assemble into a macroscopic spiral object during the filtration process, as shown in the photograph in Fig.\,\ref{fig:fig2}a.

We performed CD spectroscopy measurements using a standard CD spectrometer in a wavelength range of $200-900$\,nm with a customized 3D-printed cuvette; see Methods and Supplementary Fig.\,3 for more details. One must carefully eliminate effects of linear dichroism and linear birefringence to assess the true CD signal when the sample under study has structural anisotropy~\cite{ShindoEtAl1990BORB,AlbanoEtAl2020CR}. Hence, we adopted a four-configuration measurement approach~\cite{MertenEtAl2008AS,HirschmannEtAl2021SM}; see Methods. Figure~\ref{fig:fig2}b illustrates this approach, showing measured CD spectra for a highly aligned CNT film under four configurations together with their average. As expected for an aligned film with a racemic mixture of CNTs, there is nearly zero ($<0.5\,$mdeg/nm) CD observed. A randomly oriented CNT film and an aligned CNT film prepared using a shear force alignment technique (see Methods) also show negligible ($<0.1$\,mdeg/nm) CD signals (Supplementary Figs.\,1 and 4).

We then employed this four-configuration approach to measure CD for a twisted CNT film prepared through mechanical rotation. In addition, we performed spectroscopic ellipsometry measurements and obtained all 16 transmission Mueller matrix elements at wavelengths from 190 to 1700\,nm. The CD spectrum was obtained based on the differential decomposition of measured Mueller matrix~\cite{ArwinEtAl2021AS} (see Methods and Supplementary Fig.\,5). Figure~\ref{fig:fig2}c shows excellent agreement between the spectrum obtained through the four-configuration CD measurements and that from the ellipsometry measurements. Further, we were able to tune the strength of the CD signal by controlling the time period of applied rotational motion. Figure~\ref{fig:fig2}d shows CD spectra for different rotation times. The maximum ellipticity obtained was 19~$\pm$~1\,mdeg/nm at 263\,nm for 70\,s rotation. Supplementary Fig.\,6 shows the ellipticity at two wavelengths as a function of rotation time. 

\begin{figure}[hbt]
    \centering
    \includegraphics[width=1.0\textwidth]{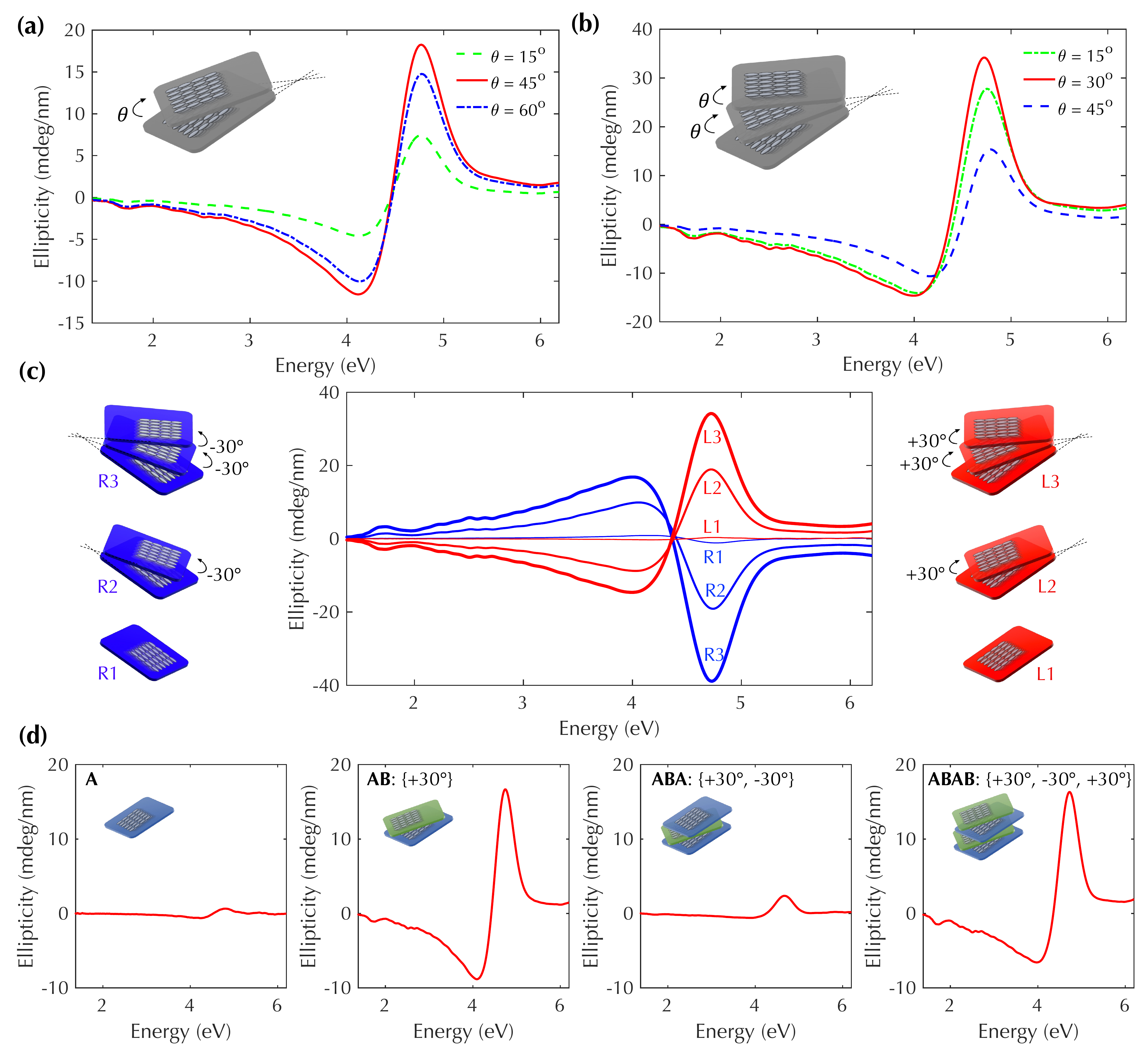}
    \caption{\textbf{Controlled chirality in twist-stacking-produced multiple-layered architectures of aligned CNTs.} (a)~CD spectra for a twisted two-layer CNT stack at twist angles of 15$^{\circ}$, 45$^{\circ}$, and 60$^{\circ}$. (b)~CD spectra for a twisted three-layer CNT stack with two equal twist angles, $\theta =$ 15$^{\circ}$, 30$^{\circ}$, and 45$^{\circ}$. (c)~CD spectra for left- and right-handed twisted three-layer stacks. All twist angles in left-handed stacks expressed as positive angles ($+30^{\circ}$), while those in right-handed stacks are expressed as negative angles ($-30^{\circ}$). (d)~CD spectra for CNT architectures in the \textbf{A}, \textbf{AB}, \textbf{ABA}, and \textbf{ABAB} twist configurations. All twist angles have the same absolute value of $30^{\circ}$.}
    \label{fig:fig3}
\end{figure}

The second method we developed for creating wafer-scale chiral CNT architectures is based on twist-stacking of multiple aligned CNT films prepared via CVF. Figure~\ref{fig:fig3}a illustrates a twisted two-layer stack. After first transferring an aligned CNT film onto a substrate, we then transferred the other aligned film on top of the first film with a twist angle $\theta$ between two alignment directions. The largest ellipticity occurs when the twist angle is $45^{\circ}$. We then repeated the same procedure for placing a third film on the second film with the same twist angle $\theta$ (Fig.\,\ref{fig:fig3}b); see Methods and Supplementary Fig.\,7 for more details. In the twisted three-layer stack, the largest ellipticity occurs when the angle is 30$^{\circ}$.

The sign of ellipticity can be controlled by the twist direction. Figure~\ref{fig:fig3}c displays CD spectra for twisted three-layer stacks with a twist angle of $30^{\circ}$ in left-handed and right-handed manners, respectively. We use positive (negative) twist angles for left-handed (right-handed) structures, and the ellipticity has opposite signs at a given photon energy between the left- and right-handed structures. It should be noted that the experimentally recorded ellipticity, 40~$\pm$~1\,mdeg/nm, in the three-layer right-handed stack at 4.77\,eV (or 260\,nm) is the highest ever obtained in the DUV range; see Supplementary Table\,1 for comparison with other chiral platforms and structures.

Furthermore, we can switch on and off the CD signal by stacking aligned CNT films with twist angles alternating between positive and negative values. Figure~\ref{fig:fig3}d shows a series of four spectra, corresponding to one-layer, two-layer, three-layer, and four-layer samples. The one-layer sample shows nearly negligible CD, and the two-layer sample with a twist angle of 30$^\circ$ shows substantial CD, as shown before.  We refer to the alignment directions of the first and second layers in the twisted two-layer stack as \textbf{A} and \textbf{B}, respectively. When we stack a third layer with direction \textbf{A}, the obtained architecture corresponds to an \textbf{ABA} configuration in which CD is strongly suppressed, i.e., CD is switched off. When a fourth layer is stacked to construct a configuration of \textbf{ABAB}, CD is restored. 

\begin{figure}[hbt]
    \centering
    \includegraphics[width=1.0\textwidth]{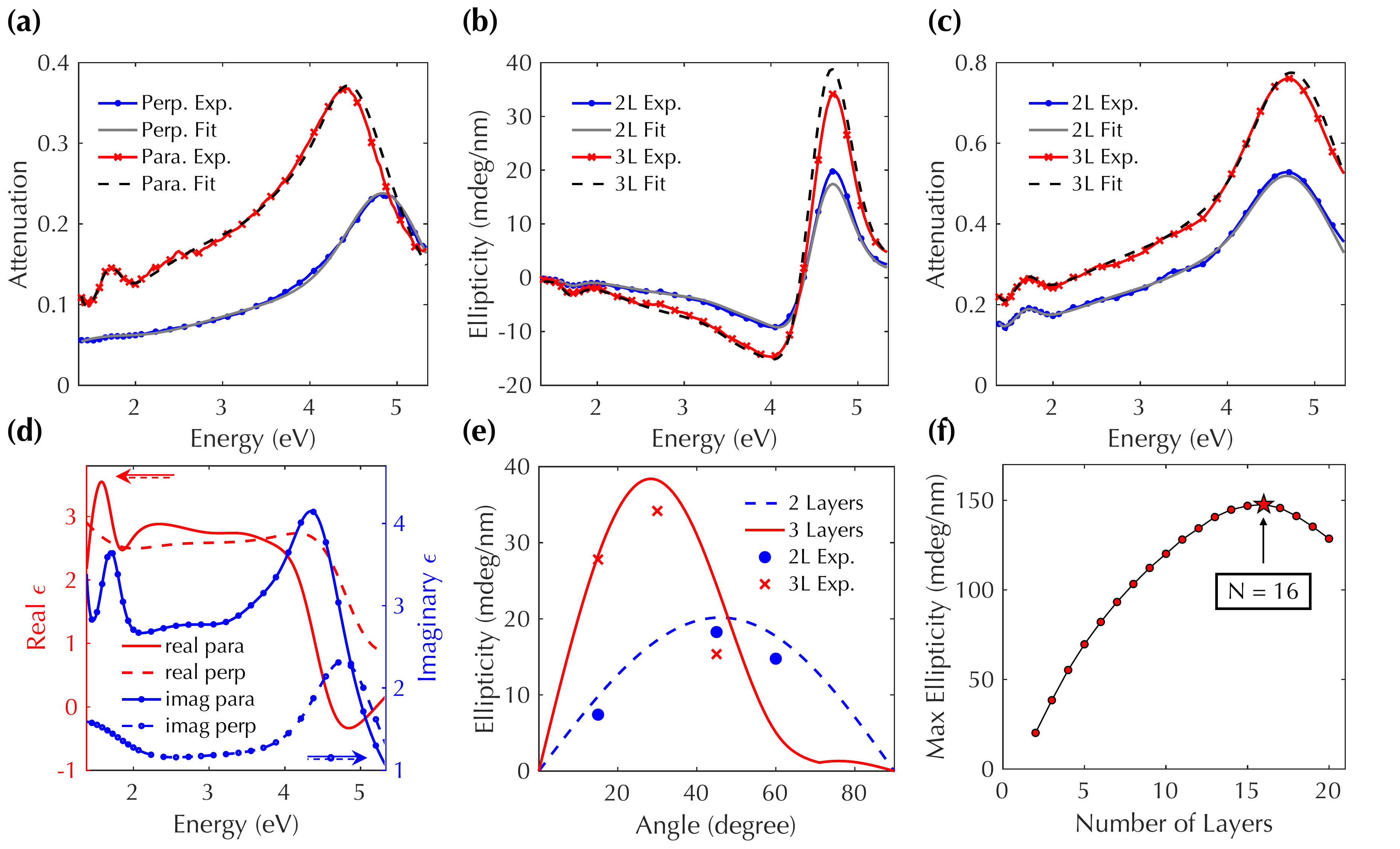}
    \caption{\textbf{Simulations and analysis of CNT-based synthetic chiral matter using the transfer matrix method.} Experimental (a)~linearly polarized absorption spectra for a highly aligned film, (b)~CD spectra, and (c)~unpolarized absorption spectra for twisted two-layer and three-layer stacks with a twist angle of 30$^{\circ}$, together with fitting curves. (d)~Extracted complex-valued anisotropic dielectric functions parallel and perpendicular to the nanotube axis. (e)~Calculated ellipticity as a function of twist angle for twisted two-layer and three-layer stacks, which agrees with experimental results. (f)~Calculated ellipticity as a function of layer number in twisted stacks.}
    \label{fig:fig4}
\end{figure}

The chiral architectures prepared through the twist-stacking approach described above have well-defined structures that are convenient for analysis using the transfer matrix method. We assumed that the parallel and perpendicular dielectric functions of aligned CNT films can be modeled as a summation of Voigt functions with multiple fitting parameters (see Methods). We then simultaneously fit six experimentally measured spectra: linear-polarization absorption spectra for a highly aligned CNT film, and CD spectra and unpolarized spectra of twisted two-layer and three-layer stacks with a twist angle of 30$^{\circ}$. With such simultaneous fitting, we were able to uniquely determine the fitting parameters and anisotropic dielectric functions. Figures~\ref{fig:fig4}a--c show fitting and experimental spectra, which all demonstrate excellent agreement. Figure~\ref{fig:fig4}d displays the extracted anisotropic complex-valued dielectric functions of the single-layer aligned CNT film. 

The obtained dielectric functions and the developed transfer matrix method allow us to understand how the CD signal changes as a function of structure parameters. Figure~\ref{fig:fig4}e shows the calculated ellipticity at 260\,nm for twisted two-layer and three-layer stacks as a function of twist angle. The largest ellipticity occurs at the twist angle 46~$\pm$~1$^{\circ}$ (28~$\pm$~1$^{\circ}$) for a two-layer (three-layer) stack, which agree well with experimental results in Figs\,\ref{fig:fig3}a and \ref{fig:fig3}b. We also analyzed the twisted two-layer stack using the Jones calculus. The calculation results not only agree with those obtained from the transfer matrix method, but also demonstrate that the anisotropic phase response in aligned CNTs is responsible for the observed CD signals in twisted stacks; see Methods and Supplementary Fig.\,8 for more details. We further calculated, for an $N$-layer stack ($N = 2-20$), the optimal twist angle, i.e., the angle that produces the largest ellipticity, assuming that the twist angle is constant throughout the $N$-layer stack. Figure~\ref{fig:fig4}f shows the largest ellipticity as a function of $N$, which exhibits a peak value of 150\,mdeg/nm at $N=16$. The decreasing ellipticity with increasing $N$ for $N>16$ can be understood as a result of nearly constant CD and linearly increasing stack thickness with increasing $N$. Supplementary Figure\,9 also displays the optimal twist angle as a function of $N$. These analyses offer promising insights into the possibility of creating synthetic CNT-based twisted architectures with stronger chiroptical responses. 

Our observation of giant CD signals in the DUV range is noteworthy because of unique properties of this band. For example, most bio- and chemical molecules have large and dispersive refractive indices in the DUV because of their electronic transitions. Thus, high-performance DUV chiral sensors will not only enable ultrasensitive detection of molecules but also can differentiate enantiomers~\cite{MohammadiEtAl2018P,YooEtAl2019N,HuEtAl2019P,WarningEtAl2021N,BothEtAl2022N}, which can have substantial impacts on pharmaceutical and biomedical science~\cite{NguyenEtAl2006IJBSI,SmithEtAl2009TS}. Furthermore, DUV photons in the sunlight are strongly absorbed by the atmosphere so that their propagation is solar-blind and not affected by the background sun radiation. Hence, incorporating chiral matter systems to construct intrinsically chiral-light-sensitive solar-blind detectors can boost the security and anti-interference capabilities of detectors, which enable new functionalities in DUV applications such as missile detection, environment monitoring, non-line-of-sight communication, and astrophysics~\cite{UlmerEtAl1995,RazeghiEtAl2002PI}.

Furthermore, in contrast to chiral metamaterials that require sophisticated nanofabrication processes to engineer conventional metallic and dielectric materials, the demonstrated synthetic chiral matter consisting of ordered CNTs can be prepared through simple and scalable procedures. Especially in the DUV range, the CNT-based chiral platform is advantageous over chiral metamaterials because the available DUV metallic and dielectric materials are limited to only a few candidates, including magnesium~\cite{JeongEtAl2016CC}, aluminum~\cite{HuangEtAl2022OE,DavisEtAl2020NL,LeiteEtAl2022NL}, and titanium dioxide~\cite{SarkarEtAl2019NL}, and the ultrasmall feature sizes and sophisticated geometries of metamaterials make their fabrication extremely challenging. In addition, the demonstrated synthetic chiral matter is fundamentally different from ordinary metamaterials because CNTs are not simple metallic or dielectric rods. Rather, CNTs are quantum wires with extreme quantum confinement. The quantized energy levels, or subbands, in CNTs manifest themselves as peaks in optical spectra, and the fact that our observed wavelength-dependent giant CD signals are due to one-dimensional van Hove singularities. The produced architectures represent an entirely new type of a chiral platform, where millions of densely ordered quantum wires are assembled into a spiral castle. These structures can potentially reveal new phenomena, including non-optical phenomena, and open up new opportunities for next-generation electronic, photonic, and optoelectronic devices.

\newpage
\section*{Methods}

\smallskip
\noindent\textbf{Standard CVF} -- The original CNT powder was made using the arc-discharge approach and was purchased from Carbon Solutions, Inc. with a product number P2. 8\,mg CNT powder was mixed with 20\,mL $0.5\%$ weight concentration sodium deoxycholate aqueous solution and then was dispersed under ultrasonic tip horn sonication for 45\,minutes at 21\,W output power. The sonicated suspension was further purified to remove large bundles through ultracentrifugation at 38000\,rpm for two hours. The supernatant was collected and diluted by 20 times for vacuum filtration. A 2" filtration system (MilliporeSigma) and 200-nm-pore-size filter membranes (Whatman Nuclepore Track-Etched polycarbonate hydrophilic membranes, MilliporeSigma) were used. The filtration process was performed in a well-controlled manner, consisting of multiple stages including slow filtration, rapid filtration, and drying processes. Once the polycarbonate filter membrane with CNTs deposited on top was fully dry, the obtained film was transferred onto the desired substrate through a wet transfer process. Specifically, a small droplet of water was first placed on top of the desired substrate. The CNT film was flipped to have the CNT surface to be in contact with the wet substrate and the polycarbonate surface was on top. The water between the CNT film and the substrate was then dried. The top polycarbonate layer was removed by submerging the sample into a chloroform solution. Finally, the sample was cleaned with isopropanol. The more detailed description can be found in our early work~\cite{HeetAl16NN}. 

\smallskip
\noindent\textbf{Characterization of highly aligned CNT films} -- The obtained highly aligned CNT films prepared using standard CVF were characterized using multiple techniques. The linearly polarized optical absorption spectra were measured using a UV-visible-near-infrared (UV-Vis-NIR) spectrometer (Perkin Elmer Lambda 950 UV-Vis-NIR) equipped with an automatically controlled rotating polarizer. The measurement wavelength range was $200-900$\,nm and the incident beam diameter was $2.2\,\textrm{mm}$, which was determined by a customized sample holder as described below. The microscopic structure was characterized using a scanning electron microscope (FEI TENEO). The surface morphology was characterized using an atomic force microscope (Parksystems NX20). 

\smallskip
\noindent\textbf{Mechanical-rotation-assisted CVF} -- The filtration system was mounted on top of an orbital shaker (Wincom KJ-201BD) using a 100 mL flask clamp. The overall system was rotated in a circle with a diameter $\sim22\,$mm. The rotation speed was 120\,rpm.

\smallskip
\noindent\textbf{Four-configuration CD measurement} -- CD spectra were measured using a standard CD spectrometer Jasco J-810 in the wavelength range of $200-900$\,nm. The incident beam diameter was $2.2\,$mm, which was defined by the sample holder described below. In general, the measured CD signals from solid-state samples (CD$_\textrm{measured}$) can be expressed as CD$_\textrm{measured}$ = CD$_\textrm{true}$ + 0.5(LBLD$^{'}$ + LB$^{'}$LD) + sin$\alpha$(LD$^{'}$sin($2\theta$) - LDcos($2\theta$))~\cite{HirschmannEtAl2021SM}. LD$^{'}$ and LB$^{'}$ are the linear dichroism and linear birefringence for the $\pm45^{\circ}$ reference directions with respect to $xy$ axes~\cite{ArwinEtAl2021AS}. The first term CD$_\textrm{true}$ describes the true CD in solid-state samples. The second term 0.5(LBLD$^{'}$ + LB$^{'}$LD) originates from linear optical anisotropy. The third term sin$\alpha$(LD$^{'}$sin($2\theta$) - LDcos($2\theta$)) comes the residual static birefringence of the modulator (angle $\alpha$) in the CD spectrometer and the in-plane sample rotation angle $\theta$~\cite{HirschmannEtAl2021SM}. The averaging of CD spectra with $90^{\circ}$ in-plane rotation can cancel the third term and the averaging of CD spectra with $180^{\circ}$ out-of-plane rotation can cancel the second term. 

In order to facilitate this four-configuration CD measurements with a commercial CD spectrometer, we designed and manufactured a customized sample holder cuvette using stereolithography 3D printing. Supplementary Figure\,3 displays the schematic of the designed cuvette main assembly and pocket, which were printed using an Anycubic Mono 3D printer. Two printed parts were then joined together with two bearings and two apertures with a 2.2-mm diameter fixed on both sides of the cuvette to define the light beam size. Finally, we attached copper foils on both sides to block the light passage from other portions. The pocket can rotate $360^{\circ}$. Thus, the solid CNT samples can be measured with the CNT side facing the light beam with varying in-plane rotation angles. The cuvette can be rotated by $180^{\circ}$ and the other side of substrate faces the light beam with varying in-plane rotation angles. As a result, we can obtain CD spectra under four configurations. 

\smallskip
\noindent\textbf{Aligned CNT films fabricated using shear force technique} -- To validate the four-configuration CD measurements, we also prepared an aligned CNT film using a shear force technique. Specifically, we used the CNT powder of Meijo DX 302, which was also a racemic mixture of semiconducting and metallic CNTs. We used few-walled CNTs ($1-2$ walls) with an average diameter $\sim2$\,nm, an average aspect ratio $\sim3800$, and an average length $\sim7\,\mu$m(determined using a capillary thinning extensional rheometry technique previously reported~\cite{TsentalovichEtAl2017AMI}). The CNT powder was added to chlorosulfonic acid HClSO$_3$ and mixed until homogenously and fully dissolved. The CNTs were aligned by sandwiching the solution between two glass slides, pressing the slides together to remove any air bubbles, applying a shearing force in the direction of the long axis of the slide, and separating the slides. The slides were then placed in a solution of 97 vol\% diethyl ether and 3 vol\% oleum to remove the chlorosulfonic acid. The slide was removed from the coagulant. The aligned film formed on one glass slide floated on water by submerging it into water. The floating film was transferred to a fused silica substrate by scooping the film with the substrate and drying the film afterwards~\cite{HeadrickEtAl2018AM}. Supplementary Figure\,4a confirms the strongly anisotropic absorption in the obtained film and Supplementary Fig.\,4b shows the negligible averaged CD signal measured using four-configuration measurements. 

\smallskip
\noindent\textbf{Spectroscopic ellipsometry} -- The spectra of 16 full transmission Muller matrix elements were measured using an RC2 ellipsometer from J.A. Woollam Company. The measurement was done under normal incidence. The incident beam diameter was $\sim$2.2\,mm adjusted by an aperture in a wavelength range of $190-1700$\,nm. All Muller matrix elements were normalized with respect to the first element $m_{11}$. The obtained Muller matrix is denoted as $M$. The natural logarithm of $M$, which is $L$, consists of CD as shown in the following expression

\begin{equation*} \label{eq:eq1}
    L = \textrm{ln}(M) = \textrm{ln} \begin{pmatrix}
    1 & m_{12} & m_{13} & m_{14}\\
    m_{21}  & m_{22} & m_{23} & m_{24}\\
    m_{31} & m_{32} & m_{33} & m_{34}\\
    m_{41} & m_{42} & m_{43} & m_{44}\\
    \end{pmatrix} = \begin{pmatrix}
    0 & -\textrm{LD} & -\textrm{LD}^{'} & \textrm{CD}\\
    -\textrm{LD}  & 0 & \textrm{CB} & -\textrm{LB}^{'}\\
    -\textrm{LD}' & -\textrm{CB} & 0 & \textrm{LB}\\
    \textrm{CD} & \textrm{LB}^{'} & -\textrm{LB} & 0\\
    \end{pmatrix}.
\end{equation*}

In addition, we rotated the in-plane polarization every $10^{\circ}$ for a full round and we averaged all CD spectra at various in-plane polarization angles, so that the influence of imperfect light polarization states in the ellipsometer was negligible. 

\smallskip
\noindent\textbf{Twist-stacking of aligned CNT films} - The building block for the twist-stacking method was a 14-nm thick 2" highly aligned CNT film prepared via standard CVF. The alignment direction of the film was first determined by taking optical images under a microscope to observe the groove direction on the filter membrane~\cite{KomatsuetAl20NL}. The film was then cut into smaller pieces using a sharp razor blade. After transferring the first film on a fused silica substrate, the whole sample was placed on top of a transparent protractor that was back-illuminated using an LED panel. The second film was then fixed on top of the first one with a rotation angle and transferred using the same wet transfer process. Supplementary Figure\,7 shows the photographs of prepared two-layer and three-layer stacks. Such process can be repeated for multiple layers with the same or different rotation angles.  

\smallskip
\noindent\textbf{Transfer matrix method} - A $4\times4$ transfer matrix method was used to calculate the optical response of multiple layers of anisotropic non-magnetic materials~\cite{YehEtAl1979JOS,HaoEtAl2008PR}. Here, we detail the $4\times4$ transfer-matrix method for calculating the response of anisotropic layered media under normal incidence. This method can avoid the singularity issue for calculating isotropic media under normal incidence using the generalized $4\times4$ transfer-matrix method~\cite{HaoEtAl2008PR}. Specifically, for a material with orthogonal principal axes, such as the directions parallel and perpendicular to CNT alignment, the electromagnetic waves in the $j$-th layer are expressed as

\begin{align*}
    \left(\begin{array}{cc} \vec{E_j}\\ \vec{H_j} \end{array}\right) &= \sum^{4}_{\sigma=1}E_{j,\sigma} \left(\begin{array}{cc} \vec{e_{j,\sigma}}\\ \vec{h_{j,\sigma}} \end{array}\right) \textrm{exp}(ik_{j,\sigma}z - i\omega t), \\
    \vec{h_{j,\sigma}} &= \frac{k_{j,\sigma}\hat{z}\times \vec{e_{j,\sigma}}}{\omega\mu_0}, \\
    k_{j,1} &= -k_{j,2} = k_0n_{s_j}, \\
    k_{j,3} &= -k_{j,4} = k_0n_{p_j}, \\
    \vec{e_{j,1}} &= \vec{e_{j,2}} = \textrm{sin}\theta_j\hat{y} + \textrm{cos}\theta_j\hat{x}, \\
    \vec{e_{j,3}} &= \vec{e_{j,4}} = \textrm{cos}\theta_j\hat{y} - \textrm{sin}\theta_j\hat{x}, \\
    \vec{h_{j,1}} &= -\vec{h_{j,2}} = \frac{k_0n_{s_j}}{\omega\mu_0}(-\textrm{sin}\theta_j\hat{x} + \textrm{cos}\theta_j\hat{y}), \\
    \vec{h_{j,3}} &= -\vec{h_{j,4}} = \frac{k_0n_{p_j}}{\omega\mu_0}(-\textrm{cos}\theta_j\hat{x} - \textrm{sin}\theta_j\hat{y}), \\
\end{align*}
where $n_{s_j}$ is the dielectric function along the $s_j$ axis of the $j$-th layer, $n_{p_j}$ is the dielectric function along the $p_j$ axis of the $j$-th layer, $\theta_j$ is the angle of the $s_j$ axis with respect to the $x$-axis in counterclockwise rotation, $k_0$ is vacuum wavevector, $z$ is the propagation distance along $z$-axis, and $\omega$ is angular frequency. 

The dielectric function tensor $\varepsilon_j$ of the $j$-th layer in $xy$ coordinate systems can be written as 

\begin{align*}
    \varepsilon_j = 
    \boldsymbol{R_j}\left(\begin{array}{cc}
        \varepsilon_{s_j} & 0 \\0 & \varepsilon_{p_j}
    \end{array}\right)\boldsymbol{R_j}^{-1}, \\
    \boldsymbol{R_j} = \left(\begin{array}{cc}
        \textrm{cos}\theta_j & -\textrm{sin}\theta_j \\\textrm{sin}\theta_j & \textrm{cos}\theta_j
    \end{array}\right).
\end{align*}

Applying the boundary conditions that tangential components of $\vec{E}$ and $\vec{H}$ are continuous at the boundary of each layer, we obtain

\begin{equation*}
    \left(\begin{array}{c}
    E_{j+1,1}\\E_{j+1,2}\\E_{j+1,3}\\E_{j+1,4}
    \end{array}\right)=\boldsymbol{D_{j+1}^{-1}D_jP_j}\left(\begin{array}{c}
    E_{j,1}\\E_{j,2}\\E_{j,3}\\E_{j,4}
    \end{array}\right)=\boldsymbol{M_{j+1,j}P_j}\left(\begin{array}{c}
    E_{j,1}\\E_{j,2}\\E_{j,3}\\E_{j,4}
    \end{array}\right)
\end{equation*}
    with
\begin{equation*}
    \boldsymbol{D_j}=\left(\begin{array}{cccc}
    \textrm{sin}\theta_j & \textrm{sin}\theta_j & \textrm{cos}\theta_j & \textrm{cos}\theta_j \\
    -n_{s_j}\textrm{sin}\theta_j & n_{s_j}\textrm{sin}\theta_j & -n_{p_j}\textrm{cos}\theta_j & n_{p_j}\textrm{cos}\theta_j \\
    n_{s_j}\textrm{cos}\theta_j & -n_{s_j}\textrm{cos}\theta_j & -n_{p_j}\textrm{sin}\theta_j & n_{p_j}\textrm{sin}\theta_j \\
    \textrm{cos}\theta_j & \textrm{cos}\theta_j & -\textrm{sin}\theta_j & -\textrm{sin}\theta_j \\
    \end{array}\right)
\end{equation*} 
    and
\begin{equation*}
    \boldsymbol{P_j}=\left(\begin{array}{cccc}
    e^{ik_{j,1}d_j} & 0 & 0 & 0\\
    0 & e^{ik_{j,2}d_j} & 0 & 0\\
    0 & 0 &  e^{ik_{j,3}d_j} & 0\\
    0 & 0 & 0 &  e^{ik_{j,4}d_j}
    \end{array}\right),
\end{equation*}
where $d_j$ is the thickness of the $j$-th layer. 

The amplitude of electromagnetic components in the last ($N$-th) layer can be related to the amplitude in the first layer by multiplying all matrices as  
\begin{equation*}
    \left(\begin{array}{c}
    E_{1,N}\\E_{2,N}\\E_{3,N}\\E_{4,N}
    \end{array}\right)=\boldsymbol{M_{N,N-1}P_{N-1}M_{N-1,N-2}P_{N-2}}\cdot\cdot\cdot \boldsymbol{M_{2,1}P_1}\left(\begin{array}{c}
    E_{1,1}\\E_{2,1}\\E_{3,1}\\E_{4,1}
    \end{array}\right)=\boldsymbol{Q}\left(\begin{array}{c}
    E_{1,1}\\E_{2,1}\\E_{3,1}\\E_{4,1}
    \end{array}\right).
\end{equation*}

$\boldsymbol{Q}$ is the transfer matrix of the whole system, from which we can calculate mode-dependent complex transmission and reflection coefficients. Mode indices 1, 2, 3, and 4 correspond to $s$-wave forward, $s$-wave backward, $p$-wave forward, and $p$-wave backward modes, respectively. The transmitted field $(E_{t,s_N}, E_{t,p_N})$, reflected field $(E_{r,s_N}, E_{r,p_N})$, and the incident field $(E_{i,s_N}, E_{i,p_N})$ can be related through

\begin{equation*}
    \left(\begin{array}{c}
        E_{t,s_N}\\ 0 \\E_{t,p_N}\\ 0 
    \end{array}\right)
    =Q\left(\begin{array}{c}
        E_{i,s_1}\\ E_{r,s_1} \\E_{i,p_1} \\ E_{r,p_1}
    \end{array}\right)
    =\left(\begin{array}{cccc}
        Q_{11} & Q_{12} & Q_{13} & Q_{14} \\
        Q_{21} & Q_{22} & Q_{23} & Q_{24} \\
        Q_{31} & Q_{32} & Q_{33} & Q_{34} \\
        Q_{41} & Q_{42} & Q_{43} & Q_{44} \\    
    \end{array}\right)\left(\begin{array}{c}
        E_{i,s_1}\\ E_{r,s_1} \\E_{i,p_1} \\ E_{r,p_1}
    \end{array}\right).
\end{equation*}

As a result, we can derive transmission and reflection coefficients in terms of the matrix elements of $\boldsymbol{Q}$ as

\begin{equation*}
    r_{ss} = \left.\frac{E_{r,s_1}}{E_{i,s_1}}\right\vert_{E_{i,p_1} = 0} 
    = \frac{Q_{24}Q_{41} - Q_{21}Q_{44}}{Q_{22}Q_{44} - Q_{24}Q_{42}},
\end{equation*}

\begin{equation*}
    r_{sp} = \left.\frac{E_{r,p_1}}{E_{i,s_1}}\right\vert_{E_{i,p_1} = 0} 
    = \frac{Q_{21}Q_{42} - Q_{22}Q_{41}}{Q_{22}Q_{44} - Q_{24}Q_{42}},
\end{equation*}

\begin{equation*}
    t_{ss} = \left.\frac{E_{t,s_N}}{E_{i,s_1}}\right\vert_{E_{i,p_1} = 0} = Q_{11}+\frac{Q_{12}(Q_{24}Q_{41}-Q_{21}Q_{44})+Q_{14}(Q_{21}Q_{42}-Q_{22}Q_{41})}{Q_{22}Q_{44}-Q_{24}Q_{42}}
\end{equation*}

\begin{equation*}
    t_{sp} = \left.\frac{E_{t,p_N}}{E_{i,s_1}}\right\vert_{E_{i,p_1} = 0} = Q_{31}+\frac{Q_{32}(Q_{24}Q_{41}-Q_{21}Q_{44})+Q_{34}(Q_{21}Q_{42}-Q_{22}Q_{41})}{Q_{22}Q_{44}-Q_{24}Q_{42}}\ .
\end{equation*}


\begin{equation*}
    r_{ps} = \left.\frac{E_{r,s_1}}{E_{i,p_1}}\right\vert_{E_{i,s_1} = 0} 
    = \frac{Q_{24}Q_{43} - Q_{23}Q_{44}}{Q_{22}Q_{44} - Q_{24}Q_{42}},
\end{equation*}

\begin{equation*}
    r_{pp} = \left.\frac{E_{r,p_1}}{E_{i,p_1}}\right\vert_{E_{i,s_1} = 0} 
    = \frac{Q_{23}Q_{42} - Q_{22}Q_{43}}{Q_{22}Q_{44} - Q_{24}Q_{42}},
\end{equation*}

\begin{equation*}
    t_{pp} = \left.\frac{E_{t,s_N}}{E_{i,p_1}}\right\vert_{E_{i,s_1} = 0} = Q_{33}+\frac{Q_{32}(Q_{24}Q_{43}-Q_{23}Q_{44})+Q_{34}(Q_{23}Q_{42}-Q_{22}Q_{43})}{Q_{22}Q_{44}-Q_{24}Q_{42}},
\end{equation*}

\begin{equation*}
    t_{ps} = \left.\frac{E_{t,p_N}}{E_{i,p_1}}\right\vert_{E_{i,s_1} = 0} = Q_{13}+\frac{Q_{12}(Q_{24}Q_{43}-Q_{23}Q_{44})+Q_{14}(Q_{23}Q_{42}-Q_{22}Q_{43})}{Q_{22}Q_{44}-Q_{24}Q_{42}}.
\end{equation*}

For layers with isotropic media, such as the input and output air layers, the principal axes are chosen to be $s_1 = s_N = \hat{x}$ and $p_1 = p_N = \hat{y}$. 

In addition, the dielectric functions parallel and perpendicular to the CNT alignment direction from the near-infrared to UV ranges were modeled as a summation of Voigt functions~\cite{MenesesEtAl2005JNS}

\begin{equation*}
    \varepsilon_{s,p}(\omega) = \varepsilon_{\infty,s,p} + \sum_{n=1}^{N}\left.C_{V,s,p}(\omega)\right\vert_{A_{n}, \omega_{0,n}, \gamma_{L,n}, \gamma_{G,n}}
\end{equation*}
with 
\begin{align*}
    \left.C_V(\omega)\right\vert_{A, \omega_{0}, \gamma_{L}, \gamma_{G}} &= -A\frac{\textrm{Im}(F(x - x_0 - iy) + F(x + x_0 + iy))}{\textrm{Re}(F(iy))}\\
    &+ iA \frac{\textrm{Re}(F(x - x_0 + iy) - F(x + x_0 + iy))}{\textrm{Re}(F(iy))}
\end{align*}
and 
\begin{equation*}
    x = \frac{2\sqrt{\textrm{ln}2}}{\gamma_G}\omega, x_0 = \frac{2\sqrt{\textrm{ln}2}}{\gamma_G}\omega_0, y = \frac{\gamma_L\sqrt{\textrm{ln}2}}{\gamma_G},
\end{equation*}
where $\omega$ is angular frequency, $A$ is amplitude factor, $\omega_0$ is resonance frequency, $\gamma_L$ is Lorentz linewidth, $\gamma_G$ is Gaussian linewidth, and $F$ is Faddeeva function. $N$ is selected as 5. For each polarization ($s$ or $p$), there are 21 fitting parameters, including \{$A_{n}, \omega_{0,n},\gamma_{L,n},\gamma_{G,n}$, $n=1-5$\} and $\varepsilon_{\infty}$. There are in total 42 fitting parameters for both polarizations.

\smallskip
\noindent\textbf{Jones calculus for the twisted two-layer stack} - Supplementary Figure\,8a displays the schematic of a twisted two-layer stack of aligned CNTs with defined coordinates. The complex-valued field transmission coefficients parallel and perpendicular to the tube axis are denoted as $t_\parallel$ and $t_\perp$, respectively. The twist angle between two layers is denoted as $\theta$. Compared to the transfer matrix method, the analysis of twisted stacks using the Jones calculus is a simplified approach without considering multiple reflections between layers. For two-layer stacks, the Jones calculus analysis is a good approximation. Specifically, the Jones matrix for the first layer of aligned CNTs can be expressed as 
\begin{align*}
    \boldsymbol{J_\textrm{CNT}} = \left(\begin{array}{cc}
        t_\parallel & 0 \\
        0 & t_\perp
    \end{array}\right)
\end{align*}
in the $x-y$ coordinate. The rotation matrix to convert the $xy$ coordinate to the $x'y'$ coordinate is
\begin{align*}
    \boldsymbol{J_\textrm{rot}} = \left(\begin{array}{cc}
        \textrm{cos}\theta & -\textrm{sin}\theta \\
        \textrm{sin}\theta & \textrm{cos}\theta
    \end{array}\right).
\end{align*}
Hence, the Jones matrix of the second layer of aligned CNTs is the same as that of the first layer in the $x'y'$ coordinate. The input circularly polarized light of different handedness in the $xy$ coordinate can be expressed as vectors
\begin{align*}
    \boldsymbol{E_\textrm{in}} = \left(\begin{array}{c}
        1 \\ \pm i
    \end{array}\right),
\end{align*}
respectively. 

As a result, the output vector in the $x'y'$ coordinate is
\begin{align*}
    \boldsymbol{E_\textrm{out}} = \left(\begin{array}{c}
        E_{x'} \\ E_{y'}
    \end{array}\right)
    =  \boldsymbol{J_\textrm{CNT}}\boldsymbol{J_\textrm{rot}}\boldsymbol{J_\textrm{CNT}}    \boldsymbol{E_\textrm{in}},
\end{align*}
and the output intensity is 
\begin{align*}
    \boldsymbol{I} = (E_{x'}^*, E_{y'}^*) \left(\begin{array}{c}
        E_{x'} \\ E_{y'}
    \end{array}\right),
\end{align*}
where $E_{x'}^*$ ($E_{y'}^*$) is the complex conjugate of $E_{x'}$ ($E_{y'}$). For the $(1,i)^{\textrm{T}}$ input, the output intensity $I_1$ is calculated as 
\begin{equation*}
\lvert t_\perp \rvert ^4 \textrm{cos}^2\theta + 2\lvert t_\parallel \rvert ^2 \lvert t_\perp \rvert ^2 \textrm{sin}^2\theta + \lvert t_\parallel \rvert ^4\textrm{cos}^2\theta + \lvert t_\perp \rvert \lvert t_\parallel \rvert (\lvert t_\perp \rvert ^2 - \lvert t_\parallel \rvert ^2)\textrm{sin}(2\theta)\textrm{sin}\phi, 
\end{equation*}
where $\phi$ is the phase difference between $t_\parallel$ and $t_\perp$. Similarly, for the $(1,-i)^{\textrm{T}}$ input, the output intensity $I_2$ is calculated as 
\begin{equation*}
    \lvert t_\perp \rvert ^4 \textrm{cos}^2\theta + 2\lvert t_\parallel \rvert ^2 \lvert t_\perp \rvert ^2 \textrm{sin}^2\theta + \lvert t_\parallel \rvert ^4\textrm{cos}^2\theta - \lvert t_\perp \rvert \lvert t_\parallel \rvert (\lvert t_\perp \rvert ^2 - \lvert t_\parallel \rvert ^2)\textrm{sin}(2\theta)\textrm{sin}\phi. 
\end{equation*}
Hence, the difference between $I_1$ and $I_2$ as a function of twist angle follows a $\textrm{sin}(2\theta)$ dependence. As shown in Supplementary Fig.\,8b, the calculated ellipticity as a function of twist angle using the transfer matrix method can be well fit with a $\textrm{sin}(2\theta)$ function. The largest difference between $I_1$ and $I_2$ occurs when $\textrm{sin}(2\theta) = 1$ and thus $\theta=45^{\circ}$, which also agrees with our experimental results. In addition, non-zero $\phi$ is required to produce the difference between $I_1$ and $I_2$, indicating that the anisotropic phase response in aligned CNTs is responsible for observed CD signals. 

\newpage
\section*{Supplementary Information}


The Supplementary Information document includes Supplementary Figures.\,S1 -- S9 and Supplementary Table\,S1. 

\section*{Acknowledgments}

We thank Bruce Weisman and Tonya Cherkuri for useful discussions. We also thank the staff of the University of Utah Nanofab for technical assistance. J.D., M.L., and W.G.\ acknowledge support from the University of Utah startup fund. J.F.\ and W.G.\ acknowledge support from NSF through Grant No.\ 2230727. J.D.\ and J.K.\ acknowledge support from the Robert A.\ Welch Foundation through Grant No.\ C-1509 and the Air Force Office of Scientific Research through Grant No.\ FA9550-22-1-0382. O.D.\ and M.P.\ acknowledge support from the Department of Energy through Grant No.\ DE-AR0001015 (Advanced Research Projects Agency - Energy). K.Y.\ acknowledges support from JSPS KAKENHI through Grant No.\ JP22H05469, JST CREST through Grant No.\ JPMJCR17I5, and PIRE Program, Japan through Grant No.\ JPJSJRP20221202. R.S.\ acknowledge support from JSPS KAKENHI through Grant Nos.\ JP18H01810 and JP22H00283.

\section*{Declarations}

\begin{itemize}
\item \textbf{Conflict of interest}: The authors declare no competing interests.
\item \textbf{Data availability}: All data that support the findings of this study are available within the paper and the Supplementary Information and are available from the corresponding author upon request.
\item \textbf{Code availability}: All the relevant computing codes used in this study are available from the corresponding author upon reasonable request.
\item \textbf{Authors' contributions}: W.G.\ conceived the idea, designed experiments, and supervised the project. J.D.\ performed the experiments with the help of A.B.\ and under the support and guidance of J.K.\ and W.G. M.L.\ conducted theoretical modeling and calculations with the help of J.F.\ and under the guidance of W.G. O.D.\ and Y.Y.\ helped with the preparation of samples under the support of M.P.\ and K.Y., respectively. N.H.\ performed Mueller matrix spectroscopic ellipsometry measurements. R.S.\ contributed to the explanation of experimental observations.
\end{itemize}

\end{document}